\documentclass{article}

\usepackage{arxiv}

\usepackage[utf8]{inputenc} % allow utf-8 input
\usepackage[T1]{fontenc}    % use 8-bit T1 fonts
\usepackage{hyperref}       % hyperlinks
\usepackage{url}            % simple URL typesetting
\usepackage{booktabs}       % professional-quality tables
\usepackage{amsfonts}       % blackboard math symbols
\usepackage{nicefrac}       % compact symbols for 1/2, etc.
\usepackage{microtype}      % microtypography
\usepackage{lipsum}
\usepackage{tabularx}

\usepackage{graphicx}
\usepackage{subcaption}
\usepackage{adjustbox}
\usepackage{multirow}

\def\BibTeX{{\rm B\kern-.05em{\sc i\kern-.025em b}\kern-.08emT\kern-.1667em\lower.7ex\hbox{E}\kern-.125emX}}

\title{Leveraging Mobile Phone Data for Migration Flows \thanks{To appear as a book chapter in "Data Science for Migration and Mobility Studies" edited by Dr. Emre Eren Korkmaz, Dr. Albert Ali Salah.}}

\author{ 
  Massimiliano Luca \\ 
  Faculty of Computer Science, \\ 
  Free University of Bolzano \\
  Mobile and Social Computing Lab \\ 
  Bruno Kessler Foundation \\ 
  \texttt{mluca@fbk.eu} 

  \And
  
  Gianni Barlacchi \thanks{Work done prior joining Amazon} \\ 
   Amazon Alexa \\ 
   \texttt{gianni.barlacchi@gmail.com}
   
   \And 
  
  Nuria Oliver \\
  ELLIS Unit Alicante Foundation \\ 
  Data-Pop Alliance \\
  \texttt{nuria@alum.mit.edu}
   
  \And 
  
  Bruno Lepri \\ 
  Mobile and Social Computing Lab  \\ 
  Bruno Kessler Foundation \\ 
  Data-Pop Alliance \\
  \texttt{lepri@fbk.eu}

}

\begin{document}
\maketitle

\begin{abstract}
	Statistics on migration flows are often derived from census data, which suffer from intrinsic limitations, including costs and infrequent sampling. When censuses are used, there is typically a time gap – up to a few years – between the data collection process and the computation and publication of relevant statistics. This gap is a significant drawback for the analysis of a phenomenon that is continuously and rapidly changing. Alternative data sources, such as surveys and field observations, also suffer from reliability, costs, and scale limitations. The ubiquity of mobile phones enables an accurate and efficient collection of up-to-date data related to migration. Indeed, passively collected data by the mobile network infrastructure via aggregated, pseudonymized Call Detail Records (CDRs) is of great value to understand human migrations. Through the analysis of mobile phone data, we can shed light on the mobility patterns of migrants, detect spontaneous settlements and understand the daily habits, levels of integration, and human connections of such vulnerable social groups. 
This Chapter discusses the importance of leveraging mobile phone data as an alternative data source to gather precious and previously unavailable insights on various aspects of migration. Also, we highlight pending challenges that would need to be addressed before we can effectively benefit from the availability of mobile phone data to help make better decisions that would ultimately improve millions of people's lives.
\end{abstract}

\section{Introduction}
As reported by the United Nations, the number of migrants worldwide is constantly growing with an estimation of 280 million migrants in 2020\footnote{\url{https://migrationdataportal.org}}. Climate change, wars and economic distress are some of the reasons behind these increasing migratory flows.
Indeed, according to the International Organization for Migration (IOM)\footnote{\url{https://www.iom.int/}}, migration is recognized as one of the critical issues of the $21^{st}$ century, posing fundamental challenges to governments in many regions of the world. For instance, policies are needed to guarantee access to healthcare, education, jobs and public services, social integration, and many other aspects related to migration, such as infrastructure and urban planning, resource allocation, border security, etc. To implement such policies, having up-to-date data about migratory movements is of paramount importance. However, traditional data sources, \emph{e.g.} official statistics, cannot offer this type of information due to intrinsic limitations, such as low sampling frequency given that migrations may change rapidly \cite{crawley2016unpacking, pradhan2004population}. Moreover, official statistics are usually published after significant time gaps of several years (or even decades) due to the complexity of the data collection and data preparation process and high costs. Thus, alternative data sources are urgently needed.

In this context, mobile phones have become ubiquitous both in developed and developing economies. As reported by the International Telecommunication Union (ITU) and the GSMA \cite{itu_2020, gsma_2020}, in 2019 there were more than 5.2 billion unique subscribers (67\% of the world population) with eight billion active SIM cards, therefore accounting for a penetration rate with respect to the population of 103\%. While penetration rates are larger in developed economies, small island developing states, least developed countries and landlocked developing countries are also experiencing rapid growths in mobile phone adoption \cite{itu_2020}. 

In recent years, passively collected, anonymized/pseudonymized, aggregated mobile network data has been leveraged to study different types of migration (\emph{e.g.}, refugees \cite{salah2019guide}, climate migration \cite{Pastor_Escuredo_2014}, labour migration \cite{bruckschen2019refugees}, and others \cite{deville2014dynamic}) and to understand daily habits, levels of integration (\emph{e.g.,} \cite{bakker2019measuring, boy2019towards, alfeo2019using}), access to education (\emph{e.g.,} \cite{mamei2019improve}) and to health services ( \emph{e.g., }\cite{altuncu2019optimizing}), and other aspects of everyday life of  migrants (\emph{e.g., }\cite{salah2019guide}). 

 Mobile phone data also have some limitations to be considered, including biases in the data (\emph{e.g.}, certain social groups might be underrepresented, such as women, children and the elderly), ethical challenges, and privacy and regulatory problems. These issues should be carefully taken into account when dealing with mobile phone data. We introduce some of these limitations later in this Chapter, while ethical and legal challenges are discussed in details in Chapter 15.

This Chapter discusses the importance of leveraging passively collected mobile network data as an alternative data source to gather precious and previously unavailable insights on migration flows. First, Section \ref{sec:mob_phone} describes what CDRs are. In Section \ref{sec:mp_migration}, we show how to use this data source to monitor different aspects and types of migration such as refugees, climate migrants, labour migrants. In Section \ref{sec:challenges} we discuss a few key limitations that may play a crucial role in capturing indicators about migration and migrants (\emph{e.g.}, technical challenges, biases, ethical issues, regulatory and financial issues and others). Finally, in Section \ref{sec:conclusion} we derive some conclusions. 

\section{Mobile Phone Data: Call Detail Records}
\label{sec:mob_phone}
This Section describes the most commonly used type of passively collected mobile network data: Call Detail Records (CDRs). CDRs are collected by telecommunication companies for billing purposes and contain information about the customers' interactions with the mobile network. 

CDRs are generated in the Base Transmission Stations (also referred to as antennas or cell towers) every time a phone makes/receives a phone call or sends/receives a short message (SMS).
While CDRs contain many fields, the most commonly used fields for research purposes are the pseudonymized phone numbers of the caller/callee, the timestamp of when the phone call/SMS took place, the duration of the phone call and the unique identifiers (IDs) of the cell towers to which the mobile phones of the caller/callee were connected. 
Note that if the caller (or the callee) is a customer of a different mobile operator, then the cell tower ID where it was connected is unknown. Moreover, no content is registered in the CDRs, but only information about the events are recorded. Thus, CDRs are often referred to as \emph{metadata}. 

\begin{figure}[ht]
    \centering
     \includegraphics[width=0.7\textwidth]{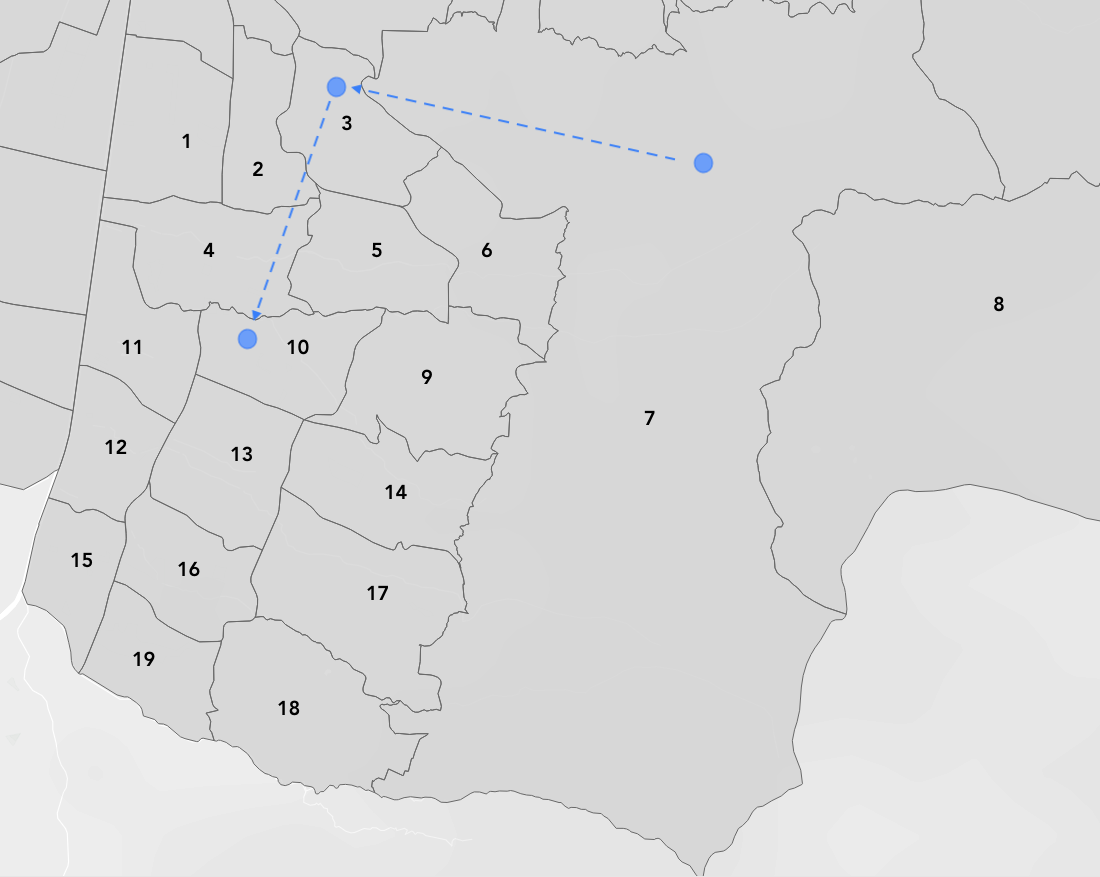}
    \caption{Example of area covered by different antennas in a cellular network. Note how the size of the area of coverage changes between urban areas (e.g. area 10) and rural areas (e.g. area 7).}
    \label{fig:antennas_medellin}
\end{figure}

In mathematical notation and for this Chapter, a CDR is a tuple $ \langle u_i, u_j, t, a_o, a_d, d \rangle $ where $u_i, u_j \in U$ ($U$ is the set of users' identifiers) and represents the encrypted identifiers of the caller and the callee; $t$ is the timestamp of when the call took place, or the SMS was sent/received; $a_o, a_d \in A$ are the identifiers of the antennas pinged at time of the event. The former ($a_o$) is the cell phone tower handling the outgoing communication. The latter ($a_d$) is the receiving tower, and $d$ is the duration of the communication which only applies to phone calls. The two antennas are known only if both caller/callee are customers of the same mobile operator. Otherwise, only one of the antennas is known. By knowing the location of the antennas and their respective areas of coverage, one can roughly estimate the position of $u_i$ and $u_j$ at an antenna level. In rural areas, the location estimation is typically within an area of a few kilometres. In urban areas, where there is a larger density of cell towers, the location estimation is within an area of around 100m by 100m. 
An example of such differences is shown in Figure \ref{fig:antennas_medellin}. The CDRs are stripped from all personal information to preserve the customers' privacy, and the customers' identifiers are pseudonymized. Table \ref{tab:cdr_example} shows an example of three CDRs of customer $u_1$ and Figure \ref{fig:antennas_medellin} shows its inferred mobility.

\begin{table}[!htb]
    \begin{tabularx}{\textwidth}{c c c c c}
        \toprule
        \textbf{Caller}     & \textbf{Callee}     & \textbf{Origin}     &   \textbf{Destination}    &   \textbf{Timestamp}    \\ \midrule
        $u_1$   &   $u_{3652}$   &   7   &   8   &   01-01-2020 11:50:32 \\
        $u_1$   &   $u_{1713}$   &   3   &   7   &   01-01-2020 14:07:24 \\
        $u_1$   &   $u_{3131}$   &   10   &   ?   &   01-01-2020 18:11:37 \\ \bottomrule
    \end{tabularx}
    
    \caption{Example of CDR for user $u_1$ for a specific date. For instance, the first line of the table shows that $u_1$ called $u_{3652}$ at 11:50:32 on the $1^{st}$ of January 2020. Moreover, we know that $u_1$ was connected to an antenna with the identifier $7$ when they made the call. Similarly, $u_{3652} $ was connected to antenna 8. Note how the last record does not have information for the antenna where the callee was connected to as $u_{3131}$ is a customer from another mobile network.}
    \label{tab:cdr_example}
\end{table}    

Although mobile phone data are not typically publicly available for privacy reasons \cite{de2018privacy}, the community has made some efforts to enable research with such data. For instance, several telecommunication companies launched data-sharing initiatives, e.g., data challenges, to release large scale CDRs for research upon requests, such as the Telecom Italia's Big Data Challenge \cite{barlacchi2015multi}. Along a similar line, the EU-funded research infrastructure SoBigData \cite{grossi2018data} opens regular calls for project proposals on datasets of various types, including mobility data \cite{andrienko2020so}. Currently, the SoBigData catalogue includes CDRs from Tuscany, Italy, and Rome, Italy.

Multiple open-source Python libraries may help to process mobile phone data and, more in general, mobility data. GeoPandas is a generic library useful to deal with geospatial data in pandas. It allows to load of different data sources (e.g., Shapefiles, geojson, and others) into a Pandas-like data structure and enables geographical operations such as spatial joins. Some libraries are more specific for human mobility. Examples of such libraries are MovingPandas \cite{graser2019movingpandas}, and scikit-mobility \cite{pappalardo2019scikit}.

Although some access limitations, the ubiquity of mobile devices has enabled the analysis of CDRs to quantify and model large-scale human mobility: 97\% of the world's population lives in an area covered by mobile cellular signal \cite{itu_2020}, and 90\% of people have access to a mobile phone \cite{poushter2016smartphone}. Thus, researchers and policymakers have an unprecedented opportunity to investigate both \emph{individual} and \emph{aggregated} mobility patterns \cite{gonzalez2008understanding, schneider2013unravelling, blondel2015survey, luca2020deep}.

\section {Application of Mobile Phone Data for Migration}
\label{sec:mp_migration}
Mobile phone data represent a unique opportunity to investigate migration. Researchers, for example, may use such data to support policymakers, organizations and authorities in taking data-driven decisions related to migration \cite{hughes2016inferring}. In the past, mobile network data have been used to solve a variety of societal challenges \cite{blondel2015survey}, including modeling migration, assessing the migrants' well-being (\emph{e.g.,} \cite{sirbu2020human, un_migration_agency_bd_migration, unhcr_connect_refugees, salah2019guide, lai2019exploring}) and estimating population displacements (\emph{e.g.,} \cite{deville2014dynamic, tatem2017worldpop, pastor2019call, koebe2020better}). This Section describes how mobile phone data have been used to monitor different types of migration.

\subsection{Refugees}
According to the United Nations High Commissioner for Refugees (UNHCR) , \emph{``refugees are people who have fled war, violence, conflict, or persecution and have crossed an international border to find safety in another country"} \cite{urd}. In recent years, the mobility and behaviours of refugees have been studied via the analysis of CDRs.  
The more prominent series of research studies related to refugees and mobile phone data was introduced with the Data for Refugees (D4R) Challenge \cite{salah2019guide}. For D4R, Türk Telekom shared a one-year CDR dataset to allow researchers, practitioners, and activists to tackle urgent social problems that emerged in Turkey after a drastic increase in Syrian refugees since the 2011 civil war. The data have been analyzed to estimate the refugees' well-being and gained insights on the refugees' living conditions (\emph{e.g.}, health, education and segregation). Differently from standard CDRs, in D4R, users are flagged as refugees when one of the following conditions holds: \emph{i)} having ID numbers given to refugees and foreigners in Turkey; \emph{ii)} being registered with Syrian passports; and/or \emph{iii)} using special tariffs reserved for refugees. It is not guaranteed to include only and exclusively refugees in the categories mentioned above, which serves as a layer of protection \cite{salah2018data}.  Multiple studies based on the analysis of this dataset have suggested ways to measure the social and economic integration of refugees (\emph{e.g.,} \cite{bakker2019measuring, boy2019towards, andris2019built, bertoli2019integration}). In particular, Bakker \emph{et al.}  \cite{bakker2019measuring} defined three metrics to measure social, spatial and economic integration. To succeeded, the authors integrated CDRs with the arriving and departing visitors by district provided by the Ministry of Culture and Tourism and with the votes per polling station during 2015 and 2018 general elections from the Ballot Result Sharing System of the Turkish Supreme Electoral Council. Social integration was measured with individual CDRs and it was defined as the ratio of calls from a minority group to a majority group and the total number of calls from minority groups (\emph{e.g.}, regardless of the group of the callee). Regarding the spatial integration, Bakker \emph{et al.} suggested using CDRs to estimate the number of people of a given group in a specific area and used the Gini coefficient to measure the presence of different groups. In this case, higher values represent more diversity and, therefore, more integration. Individual CDRs were also used to compute the probability that a minority group shares an area with a person of a majority group. Finally, the authors suggested measuring economic integration by calculating mobility similarity matrices for some key timestamps (\emph{e.g.}, weekends, evening, working hours) and used the Frobenius norm to measure the employment score.
According to the authors, refugees in Istanbul were more integrated in terms of spatial integration than those living in Southeastern Anatolia. However, there was a higher and positive correlation of spatial, social, and economic integration in Southeastern Anatolia. Finally, the authors showed that the presence and integration of refugees had an impact on political outcomes. For instance, their study showed that social integration was correlated with an increase of votes for pro-refugees parties. Economic integration was instead correlated with the opposite. 
Boy \emph{et al.} \cite{boy2019towards} measured the segregation (\emph{i.e.}, an imposed restriction on the interaction between people that are considered to be different \cite{freeman1978segregation}), isolation (\emph{i.e.}, to which extent different groups share residential areas \cite{massey1988dimensions}), homophily (\emph{i.e.}, the tendency of people to associate with groups which they share similar traits with \cite{currarini2016simple}), and they used such metrics as a proxy for the social integration of refugees. The authors used individual CDRs and aggregated mobility matrices (\emph{e.g.}, origin-destination matrices derived from CDRs) to look at the evolution of mobility and communication patterns of refugees. 

Another study that aims to investigate the integration of refugees is described in \cite{alfeo2019using}. Alfeo \emph{et al.}  computed the similarity between locals and refugees using their mobility and communication patterns as a proxy for social integration. More precisely, they defined four metrics to measure social integration: (i) district attractiveness, (ii) residential inclusion by district, (iii) refugees' interaction level, and (iv) refugees' mobility similarity. The authors found that mobility similarity is positively correlated with the refugees' interaction level, showing that sharing urban spaces among locals and refugees can improve social integration. To perform their analyses, the authors had to infer the home location of the individuals involved in the study. To do so, they used individual CDRs, and some heuristics like the ones introduced in \cite{pappalardo2020individual}. A similar study by Andris \emph{et al.} showed that the presence of certain types of amenities such as community centres, social centres, schools, and places of worship plays a crucial role in the integration of refugees \cite{andris2019built}. In this study, CDRs analyzed at an individual level are integrated with Points of Interest (POIs) derived from the Humanitarian Open Street Map \footnote{\url{https://www.hotosm.org/}}. Moreover, to measure the distances between POIs and the refugee camps, the authors used Turkish refugee campsites sourced from the UNHCR and the Regional Information Management Working Group-Europe.
Other works integrated alternative data sources to measure the refugees' social integration. This is the case of \cite{bertoli2019integration} where Bertoli \emph{et al.} relied on CDRs, the Global Dataset of Events, Language and Tone (GDELT), and housing data. Hu \emph{et al.} \cite{hu2019quantified} used both CDRs and Points-Of-Interest (POIs); while Bozcaga \emph{et al.} \cite{bozcaga2019syrian} combined mobile phone data with official statistics from different sources (\emph{e.g.}, Ministry of Interior, Turkish Census and other government statistics). Marquez \emph{et al.} \cite{marquez2019segregation} also used data from Twitter and \cite{kilicc2019use} employed multiple official statistics to explore several factors of the Syrian refugee crisis (\emph{e.g.}, home and work location of the refugees, meaningful places for the refugees, and others).  Saa \emph{et al.} combined surveys with mobile phone data to measure refugees' integration \cite{saa2020looking}. The authors used a modified version of the gravity model to contrast the mobility patterns and economic integration of people that left Venezuela to move to Colombia and of Colombians who used to live in Venezuela and went back to Colombia in the context of the recent migration crisis across South America. Using aggregated mobile phone data --combined with the monthly household survey of DANE (the Colombian statistical bureau)-- the authors showed that the drivers behind the mobility patterns and destination choices are different.

More recently, mobile phone data have been used as a source for a gravity model to understand better the factors governing the refugees' mobility (\emph{e.g.}, changes of income, propensity to leave camps, transit through Turkey and agricultural business cycles, and others) \cite{beine2021gravity}. In this study, CDRs were used to understand if mobility can be modelled with a gravity model \cite{barbosa2018human}. Mobile phone data were integrated with the Gross Domestic Product (GDP) provided by the Turkish Statistical Institute to measure income levels, and with data from the Directorate General of Migration Management in Turkey to have information about the propensity to leave camps.

Mobile phone data, integrated with additional data sources, are also employed to measure the access of refugees to healthcare services  \cite{salman2021modeling, altuncu2019optimizing}, to education \cite{mamei2019improve} and, more in general, to investigate the effect of displacement on the mobility and communication patterns \cite{beine2019refugee} \cite{frydenlund2019characterizing}.

Human flows can also be used to estimate the spread of infectious diseases. Bosetti \emph{et al.} \cite{bosetti2020heterogeneity}, used mobile phone data to quantify the risk of measles outbreaks. The authors found that heterogeneity in immunity, population distribution, and human-mobility flows are critical factors that augment the possibility of having an outbreak. Conversely, social integration and vaccine campaigns were found to be winning formulas to reduce such risks.

Finally, Mancini \emph{et al.} \cite{mancini2019opportunities} discussed opportunities and risks of using mobile phone data to monitor the refugees' daily experiences. The authors investigated the role of mobile phones in the refugees' day-to-day life, their journeys (based on mobile phone and other data sources, such as surveys and interviews), their social relations, self-empowerment, education and health. Most of the studies investigated in the review are focused on Syrian refugees and multiple countries are involved, such as Germany \cite{borkert2018best}, Jordan \cite{wall2017syrian} and others (\emph{e.g.}, \cite{smets2018way, charmarkeh2013social}). 

Several observations emerge from the literature on the analysis of mobile data related to international refugees. First, mobile phone data are primarily used to investigate the refugees' socio-economic and living conditions in the host country. Second, the use of GPS information from the refugees' smartphones may pose severe security and privacy issues \cite{beduschi2017big}. Finally, mobile phone data are usually integrated with statistical data and surveys (both official and not) to have a more complete picture of the context. 

\subsection{Climate-based Migration}

Climate change is a long-term change in the average weather patterns affecting the Earth’s local, regional and global climates. Such weather patterns are exemplified by changes in the frequency and intensity of severe phenomena, such as floods, storms, and droughts. These extreme natural phenomena have a significant impact on migrations flows and mobility patterns. Examples are the rainfalls in Burkina Faso \cite{henry2004impact} and the drought in Etiopia \cite{meze2000migration}. Environmental migration is defined by the International Organization for Migration (IOM) as characterized by \emph{``people who, predominantly for reasons of sudden or progressive changes in the environment that adversely affect their lives or living conditions, are obliged to leave their habitual homes or choose to do so, either temporarily or permanently, and who move within their country or abroad"} \cite{iom_climate_migration}. 

Two examples of leveraging location data to model climate-based migration are found in \cite{sakaki2010earthquake} and \cite{wang2014quantifying} where the authors analyzed geolocated tweets to model mobility changes after an earthquake in Japan and after Hurricane Sandy, respectively. 
Regarding the studies that rely on mobile phone data, Isaacman \emph{et al.} \cite{isaacman2018modeling} analyzed CDRs to investigate environmental migration in La Guajira, Colombia. In 2014, the area was affected by a prolonged drought period that pushed the population - mainly those living in rural zones - to move to other places. 

In this case, mobile phone data at an individual level were used to detect the home location of the users before and after the period under analysis and, therefore, to estimate the number of migrants.

The authors investigated mobility on a national level. They found out that 90\% of the people on the move stayed in the region of La Guajira but moved to a different municipality. In contrast, the other 10\% moved to other districts in Colombia. The findings confirmed a general theory indicating that only short-distance moves appeared affected by climatic factors \cite{henry2004impact}. 
To model people on the move, the authors proposed a home detection algorithm and they modelled migrants using the gravity model \cite{zipf1946p} and the radiation model \cite{simini2012universal}. Finally, since weather data play a pivotal role in mobility patterns, the authors proposed a modified version of the radiation model to deal with meteorological data.

An analysis aiming to estimate evacuation, displacement, and migration based on mobile phone data was conducted by Lu \emph{et al.} in the context of the tropical cyclone Mahasen in Bangladesh \cite{lu2016detecting}. The authors showed that mobility patterns allowed estimating the effectiveness of early warning strategies based on early and mid-storm population movements. In Chittagong, the early warning accomplished the aim of motivating evacuation during appropriate times. The authors analyzed individual CDRs of millions of users covering a period between one month before and one after it. Moreover, to reduce potential noise and biases, the authors considered only the SIMs already registered with the telecommunication company before the natural disaster. To detect the effectiveness of interventions, they investigated anomalies in the cell phone towers activities. At the same time, to estimate mobility patterns, they derived mobility matrices at a tower level by aggregating the CDRs. Similarly, Moumni \emph{et al.} highlighted that during the Oaxaca earthquake, they faced a moderate increase of mobility with respect to the baseline showing anomalies in mobility patterns \cite{Oaxaca2013}.  Also, in this study, the role of CDRs at an individual level was to detect anomalies in the activities. The studies mentioned above showed that mobility networks could eventually be used to estimate people's displacement before, during, and after a natural disaster.  
Along this line, Pastor Escuredo \emph{et al.} \cite{Pastor_Escuredo_2014} used aggregated CDRs combined with additional information such as rainfall data and remote sensing imagery to characterize the activity patterns after the floods in Tabasco, Mexico. The authors found abnormal communication patterns and stated that such insights could locate damaged areas and optimize the resource allocation in the first hours after the natural disaster. The authors also highlighted that communication activity could also be used to measure the awareness of the at-risk population.

An additional study by Lu \emph{et al.} aimed at quantifying the incidence and duration of migrations in Bangladesh and long-term mobility patterns in areas stressed from a climate perspective \cite{lu2016unveiling}. In this study, the authors also highlighted the importance of integrating mobile phone data with other targeted phone-based and household-based panel surveys to characterize vulnerable, under-represented groups in the mobile data (\emph{e.g.}, women, children and the poorest).CDRs were preprocessed as in the previously discussed study \cite{lu2016detecting}: only users that were already active before the cyclone were considered. Note that users who activated a SIM less than ten days before the cyclone were not considered.

Bengtsson \emph{et al.} showed that mobile phone data could be used to estimate the displacement of the population after the Haiti earthquake of 2010 \cite{bengtsson2011improved}. The authors demonstrated the validity of their findings by comparing the obtained results with the data collected in an extensive retrospective population-based survey carried out by the United Nations. Similarly to other studies, the authors used individual CDRs to follow users for 42 days before the natural disaster and 158 days after the disaster to detect changes in outgoing and incoming mobility patterns.

Based on the case studies conducted on Haiti's disasters (2010, 2016) and in Nepal (2015), the International Migration Organization, in cooperation with other partners, released FlowKit by Flowminder\footnote{https://www.flowminder.org}. FlowKit is a software tool that facilitates mobile phone data analysis and ensures data privacy to leverage partnerships with mobile phone operators \cite{flowminder}. In particular, FlowKit allows a near real-time assessment of population displacement after natural disasters in various contexts.

A timely analysis of the data is of paramount importance, especially when dealing with environmental migration and refugees. Most --if not all-- the work carried out so far has not been performed in real-time but after the facts. Having tools to enable fast partnerships with the custodians of the data to ease the data sharing and analysis process is thus of critical importance.

Mobile phone data are also widely used to investigate changes in the communication patterns in areas affected by natural disasters \cite{Oaxaca2013, lu2016detecting}.
In general, mobile phone data have been shown to play a crucial role in understanding the displacement of people before, during, and after natural disasters, assessing the effects of early warnings for evacuation orders, and investigating long-term mobility impacts caused by climate change. They are particularly valuable when complimented and integrated with traditional data sources, such as household surveys. 

A recently published dataset that may be useful for future investigations of climate-based migrations is the Geocoded Disasters (GDIS) dataset \cite{rosvold2021gdis}. The dataset contains geographic information about almost 40 thousand locations connected to nearly 10 thousand natural disasters. Each disaster is mapped to a class (e.g., drought, extreme temperature, storm and others). Finally, it contains an identifier that allows researchers to easily integrate information from another popular dataset that records natural disasters: EM-DAT \footnote{\url{https://www.emdat.be/}}
\subsection{Labor Migration}

The International Labor Organization (ILO)\footnote{https://www.ilo.org/global/lang--en/index.htm} defines migrant workers as \emph{''all international migrants currently employed or unemployed seeking employment in their current country of residence"}.
Obtaining timely information about labour migration is an open challenge at all the stages of the migration journey \cite{women_workers}. Alternative data sources may help track migrants obtain up-to-date statistics and measure social aspects of the migrants' lives (\emph{e.g.}, safety, access to communication technologies, social integration, and others). As an example, Neubauer \emph{et al.} used Twitter geo-tagged posts to estimate work migration patterns \cite{neubauer2015volume}. 

Recently, a study on women migrant workers in the Association of South-East Asian Nations (ASEAN) countries used mobile phones as proxies for access to the internet and new technologies \cite{women_workers}. In particular, they proposed using mobile phones as a tool for measuring access to information on their rights, support services, and respond to violence.
Al{\i}{\c{s}}{\i}k \emph{et al.} used mobile phone data to investigate mobility patterns of Syrian refugees in Turkey that might be related to seasonal jobs in fields such as seasonal agriculture and seasonal tourism \cite{alicsik2019seasonal}. The authors found that, in general, for Syrians, labour is a vital mobility driver. Antalya and Mugla's regions are considered the two regions with the higher incoming mobility related to the seasonal work opportunities in the field of tourism and Rize, Ordu, Giresun, Nigde, and Malatya for agriculture. 
As already mentioned, vulnerable groups such as refugees can be difficult to track with mobile phones. Vulnerable groups may also be exploited for cheap --or even free-- labour in high-risk scenarios.

Bruckschen \emph{et al.} \cite{bruckschen2019refugees} proposed a framework to detect fine-grained socio-economic occurrences with a limited training dataset. The authors used CDRs to spot potentially undeclared employment of refugees in Turkey by analyzing seasonal migration patterns in two scenarios: hazelnut harvest in Ordu and the construction of Istanbul's airport. In their study, work-related commuting patterns suggesting potential undeclared employment emerged. 

Olivier \emph{et al.} analyzed mobile phone data, used in combination with household survey data, to measure the impact of the Venezuelan migrants in Ecuador on the labour market \cite{olivieri2020labor}. In this case, CDRs were used to associate mobile phones held by Venezuelans to a census tract that most likely contains their home location. The authors found that regions with the higher inflows of Venezuelans do not impact the labour market or employment. In contrast, low-educated Ecuadorian workers in these regions had a reduction in employment quality and earning.

In Niger, mobile phones were leveraged to measure the impact of technology accessibility for labour migrants \cite{aker2011mobiles}. This study showed that having access to ICT increases the probability of rural to urban migrations. Such patterns are primarily due to increased communication with social networks (\emph{e.g.}, information on the labour market and work opportunities). 

Conversely, Ciacci \emph{et al.} showed that having access to mobile phone signals reduces the probability of migration by members of a household \cite{ciacci2020mobile}. This is due to the positive effects of mobile phones on the labour market and access to well-being. In this paper, however, mobile phone signal access is not measured with mobile phone data but with a survey.

\subsection{Other Usages of Mobile Phone Data}

While the works mentioned above target a specific definition or aspect of migration, 
other relevant papers use mobile phone data to tackle additional migration-related phenomena, such as internal migration and mobile phones as an alternative data source to official statistics. In this Section, we summarize the most relevant among such papers. 
Deville \emph{et al.} have analyzed mobile phone data to estimate the population density at a national and seasonal scale in Portugal and France \cite{deville2014dynamic}. This paper contains important insights on how to map a population timely and, therefore, to highlight temporal shifts that may be related to some types of internal migration. Similarly, Furletti \emph{et al.} \cite{furletti2014use} used mobile phone data to observe inter-city mobility patterns. Other studies used mobile phone data as a potential alternative to national statistics in developing economies, such as Namibia \cite{lai2019exploring} and Rwanda \cite{blumenstock2012inferring}. 

The WorldPop project \cite{tatem2017worldpop} integrates mobile phone data with censuses, surveys, and other data sources to estimate the population density worldwide with a spatial granularity 100m$\times$100m cells.
Finally, Chi \emph{et al.} \cite{chi2020general} provided a general framework to detect migration events. The authors empirically validated their approach using Twitter data for the United States and mobile phone data in Rwanda. 

\section{Mobile Phone Data for Migration: Challenges and Limitations}
\label{sec:challenges}
While CDRs have the potential to enable us to overcome many of the intrinsic limitations of traditional methods (\emph{e.g.} surveys), such as infrequent sampling and high costs, they present specific challenges that would need to be addressed before this type of data can be effectively used to measure and model human migration. 
In this Section, we highlight the main limitations of leveraging mobile network data in the context of human migrations. 
These limitations are different, including technical, regulatory, financial and ethical dimensions. 
Our goal is to highlight that such limitations should be carefully considered when using this type of data for migration-related purposes.

\subsection{Technical Challenges}
There are numerous technical challenges posed by analyzing mobile network data for migration. Below is a summary of the most prominent ones.  A more detailed discussion on some of the following limitations is presented in Chapter 15.

\noindent \textbf{Geographic and Temporal Resolution:}
While CDRs might be available for a large portion of the population, their spatial granularity depends on the geographic density of antennas, which, in turn, is proportional to the density of the population served by those antennas. Thus, coverage of an antenna might range from a few kilometres in rural areas to hundred meters in urban areas. Figure \ref{fig:antennas_medellin} shows an example of regions covered by antennas in a city. As illustrated in the Figure, the difference in coverage area between the area of coverage of antenna 7 versus that of antenna 2 stands out. 

In terms of temporal resolution, CDRs are event-based, which means that the data are only generated when the phone is used (making/receiving a phone call or sending/receiving an SMS). There are no data related to the customers' behaviours when they do not interact with their devices.  Thus, the temporal granularity of the CDRs depends on the intensity of usage of the mobile phones by the customers, leading to demographic and socio-economic differences which impact the representativeness of the data, as explained below. This limitation is partially solved by the so-called eXtended Detail Records (XDRs). XDRs measure the number of bytes downloaded/uploaded at each of the antennas. Thus, they are dramatically less sparse temporally than CDRs data but have rarely been analyzed for migration purposes. 

\noindent \textbf{Representativeness of the Data:} 
Understanding the representativeness of the data is of paramount importance when modelling migrations. Mobile phone data are known to be biased: users are unevenly distributed by demography, geography, and socio-economic groups \cite{arai2016comparative, wesolowski2013impact, wesolowski2012heterogeneous}. For example, Sekara \emph{et al. }\cite{sekara2019mobile} highlighted that in Turkey, in 2018, the mobile phone subscribers penetration was the 65\% showing that the 35\% of Turkish does not have a SIM registered to their name. In other terms, we have individuals with multiple SIMs and others with no SIM. Other studies highlight that tracking peculiar portions of the population may pose significant challenges. For example, in their study concerning the spread of Malaria, Marshall \emph{et al.} highlighted the potential biases of tracking people using mobile phones in sub-Saharan Africa regions \cite{marshall2016key}. In that region, men are more likely to be mobile phone owners, phone sharing is common among rural women, and many individuals use multiple SIM cards due to non-overlapping provider coverage. Similar biases have shown to be a key challenge in tracking children in Turkey \cite{sekara2019mobile}. Other issues related to the representativeness of mobile phone data are investigated in \cite{arai2016comparative} by Arai \emph{et al.}.

\noindent \textbf{Data Integration:} 
Combining data from different sources and/or countries/regions is especially relevant when dealing with migration. Whenever people cross a border, their mobile phones switch to roaming on the infrastructures of different operators. Hence, modelling the behaviour of international migrants requires a significant economic effort for researchers who would need to buy the data or establish partnerships with multiple companies. Moreover, a substantial technical effort is necessary to integrate  CDRs from different sources, with potentially different biases, temporal and spatial resolutions. For instance, while mobile phone records in Kenya are an excellent proxy for mobility, regardless of socio-economic factors \cite{wesolowski2013impact}, mobile phone data in Rwanda are a good proxy only for the mobility of wealthy and educated men \cite{blumenstock2012divided}. In general, integrating data with very different levels of representativeness poses significant challenges.

\noindent \textbf{Timeliness of the Data:} 
When dealing with migration, an essential goal of researchers and organizations is to provide timely insights on the situation of migrants. However, real-time access to CDRs has proven very difficult. Subsequently, it is still an open challenge. Having such an access model would require institutional partnerships with telco operators and significant investments in infrastructures and mechanisms to guarantee security and privacy \cite{pastor2019call}. One relevant initiative for this is the Open Algorithms (OPAL) project \footnote{\url{https://www.opalproject.org/}}. Launched by the MIT Media Lab, Imperial College London, Orange, the World Economic Forum, Telefonica and Data-Pop Alliance, OPAL aims to provide open source tools to unlock the potential of private sector data (mainly mobile phone data) for public good purposes by “sending the code to the data” in a safe, participatory, and sustainable manner \cite{YvesOPAL, lepri2018fair}.

\subsection{Ethical, Regulatory and Financial Challenges}
In this Section, we only introduce ethical, regulatory and financial challenges that will be discussed in details in Chapter 15.

\textbf{Ethical Challenges:}
The broad adoption of mobile phones opened new regulatory and ethical challenges. As an example, with 67\% of the world population having a subscription with a mobile phone operator, it would be possible to implement massive surveillance projects by tracking such devices. Ethical challenges may have a tremendous impact particularly on communities or groups that are already vulnerable as in the case of migrants. For instance, mobile phone data may be of critical importance to help policymakers or international organizations take actions supporting migrants. However, the same data can also be used for non-social good purposes such as discrimination, stand in the way of migrants and many others. For this reason, guaranteeing the ethical usage of data is a significant challenge. 
Competitions such as D4R, as an example of data access policies, implemented a scientific committee and a project evaluation committee aiming also to evaluate the ethical aspects of the submitted research proposals \cite{salah2019introduction}. Data were made available only after the approval of such committees. Another discussion on ethical challenges using mobile phone data especially for analyzing vulnerable communities, is carried out in \cite{taylor2016no} using the Data for Development challenge based in Côte d'Ivoire as a case study. 
A perspective of ethical challenges focusing on migration can be found in \cite{vinck2019no}.

\noindent \textbf{Privacy and Regulatory Challenges:} Privacy represents a fundamental right in modern society. Even if the data shared by the operators are aggregated and de-identified, there might be ways to re-identify a user depending on aggregation and anonymization tools and procedures used to prepare the dataset \cite{de2013unique}. 

In general, to avoid privacy issues, the data are collected and shared without including names, phone numbers or identifying information. In some cases, some socio-demographic characteristics of users may be associated with CDRs. Moreover, in theory, the users involved in the studies must agree in sharing their data for specific usage. In practice, for example, in the case of refugees, people may not have full control on how their data are collected and stored by authorities or organizations \cite{vinck2019no}. 

To guarantee people's privacy, authorities started to design frameworks such as the GDPR in Europe, CCPA in California and many others to ensure privacy to everyone. 

 Companies also use such frameworks for sharing data. However, there are cases in which countries do not have frameworks to govern the release of data. This is the case of the D4D Challenge in Côte d'Ivore \cite{de2014d4d}. Unlike the other states in West Africa that were using the same data protection regularities used in France at the time of the competition, Côte d'Ivoire did not have a data protection regulation. In that case, data have been released using a non-disclosure agreement between the researchers and the telco company. The agreement, in this case, was also the only commitment to privacy and data protection \cite{taylor2016no}.

\noindent \textbf{Difficulty of Access and lack of financially sustainable models:}
 CDRs are privately held by telecommunication companies. While several mobile phone datasets have been shared with researchers and practitioners in the context of data challenges \cite{salah2019guide,  blondel2012data}, they are generally not publicly available. Even if the data is fully anonymized, aggregated and only meant to be used for public interest purposes, access to this type of data has proven difficult. Recognizing the value that different types of privately held data could have for social good, the European Commission created in 2018 a high-level expert group on \emph{Business to Government Data Sharing}\footnote{https://ec.europa.eu/digital-single-market/en/news/meetings-expert-group-business-government-data-sharing}. This group published in February of 2020 their report \cite{towards_european_stategy} which put forward several recommendations to foster an environment where privately held data could be leveraged by the public sector for public interest purposes. 
 Similarly to other privately held datasets (\emph{e.g.} financial data, data from social platforms and digital products), the difficulties related to accessing mobile data result in (i) high costs to scientists and practitioners who need to pay to access the data; and (ii) lack of reproducibility of the  works based on the analysis of such proprietary datasets. 
In addition to the barriers mentioned before, there is also a lack of financially sustainable business models to support companies in data sharing efforts for the public interest.

\section{Conclusion}
\label{sec:conclusion}
Nowadays, more people than ever before are on the move. Policymakers face tremendous issues in developing evidence-driven policies to address the challenges associated with migrations.
In this context, mobile phones are the most widely adopted piece of technology today, with more than 5.2 billion subscribers and a penetration rate of the 103\%. Hence, pseudonymized, aggregated and passively collected mobile data have the potential to help shed light on the living conditions of migrants, providing relevant insights to policymakers, international organizations and humanitarian actors. In this Chapter, we have described the most commonly used type of mobile data, together with their peculiarities and limitations. Next, we have provided an overview of how this type of data have been used to monitor the mobility of migrants and other aspects of their life, including their integration, access to services and working conditions. 
We are excited about the opportunities that mobile phone data bring to enable us to better understand this complex problem. While mobile phone data are not exempt of limitations, it is a valuable tool to complement existing data sources and contribute to achieving our vision of evidence-driven policy-making. As mobile phone penetration increases, we are hopeful for a future where everybody counts and is counted. 

\bibliographystyle{unsrt}
\bibliography{references}
 %%% Uncomment this line and comment out the ``thebibliography'' section below to use the external .bib file (using bibtex) .





% \end{thebibliography}

\end{document}